\documentclass{revtex4}
\usepackage[cp1251]{inputenc}
\usepackage[T2A]{fontenc}
\usepackage[english,russian]{babel}
\usepackage{amssymb,latexsym,amsmath,amscd}

\usepackage[dvips]{graphicx}

\begin{document}
\title{On a new application of the path integrals in polymer statistical physics}
\author{\firstname{Yu.~A.}~\surname{Budkov}}
\email{urabudkov@rambler.ru; ybudkov@hse.ru}
\affiliation{Department of applied mathematics, National Research University Higher School of Economics, Moscow, Russia}
\author{\firstname{A.~L.}~\surname{Kolesnikov}}
\affiliation{Institut f\"{u}r Nichtklassische Chemie e.V., Universit\"{a}t Leipzig, Leipzig, Germany}

\begin{abstract}
We propose a new approach based on the path integral formalism to the calculation of the probability distribution functions of quadratic quantities of the Gaussian polymer chain in $d$-dimensional space, such as the radius of gyration and potential energy in the parabolic well. In both cases we obtain the exact relations for the characteristic function and cumulants. Using the standard steepest-descent method, we evaluate the probability distribution functions in two limiting cases of the large and small values of corresponding variables.
\end{abstract}
\maketitle
\section{Introduction}
The great efforts up to now have been made to develop the methods for calculations of the probability distribution functions (PDFs) of various quantities of the Gaussian polymer chain \cite{Flory_book,Yamakawa}. One of the most famous methods is the Wang-Uhlenbeck (WU) method which is based on the averaging of the polymer chain microscopic distribution function over the all possible conformations. So the averaging yields the macroscopic distribution function which depends on an appropriate collective variable. It should be noted that within the WU method the polymer chain is usually considered as the discrete chain of monomers in the framework of the Gaussian or freely-jointed models \cite{Yamakawa}. As is well known, both models are equivalent for the polymer chain is under not so large stretching \cite{Flory_book,Khohlov_book}.

Alternative way to calculate the PDF of some polymer chain quantities consists of the consideration of the Gaussian polymer chain as a continuous curve in $d$-dimensional space. In this case the averaging over the polymer chain conformations is equivalent to the path integration over the Winer's measure \cite{Edwards,Fred}. However, to the best of our knowledge this approach \cite{Feynman,Zinn-Justin,Efimov} to the PDF calculations is not properly addressed in the literature till now. The aim of this work is to fill this gap. We shall consider the PDF calculation for two quadratic quantities of the Gaussian polymer chain in $d$-dimensional space: the radius of gyration and potential energy in the symmetric parabolic well.

\section{Example I: PDF of radius of gyration of Gaussian polymer chain in d-dimensional space}
As is well known, a radius of gyration is one of the most important collective variables which is usually used to describe the macromolecular conformational state in experiments (see, for instance, ref. \cite{Zhang}), computer simulations \cite{Heyda2013}, and theory \cite{Grosberg1,Brilliantov2,Budkov1,Budkov2,Budkov3,Budkov4,Budkov5}. In this section we demonstrate the method of PDF calculation of the radius of gyration of Gaussian polymer chain in $d$-dimensional space based on the path integral formalism.
\subsection{Statement of problem}
We start from the PDF of radius of gyration of Gaussian polymer chain in $d$-dimensional space which can be written as the following path integral
\begin{equation}
\label{eq:distr1}
P(R_{g}^2)=\int\frac{\mathcal{D}\bold{r}}{Z_{0}}\exp\left[-\frac{d}{2b^2}\int\limits_{0}^{N}d\tau\dot{\bold{r}}^2(\tau)\right]
\delta\left(R_{g}^2-\frac{1}{2N^2}\int\limits_{0}^{N}\!\!\!\int\limits_{0}^{N}d\tau_{1}d\tau_{2}\left(\bold{r}(\tau_{1})-\bold{r}(\tau_{2})\right)^2\right),
\end{equation}
where
\begin{equation}
Z_{0}=\int \mathcal{D}\bold{r}\exp\left[-\frac{d}{2b^2}\int\limits_{0}^{N}d\tau\dot{\bold{r}}^2(\tau)\right]
\end{equation}
is the normalization constant; $\dot{\bold{r}}(\tau)=\frac{d\bold{r}(\tau)}{d\tau}$; $b$ is the Kuhn length of the segment, $d\geq 1$ is the dimension of space; $N$ is the degree of polymerization.
Using the substitution $\tau=sN$, where $0\leq s\leq 1$, the PDF can be rewritten as follows
\begin{equation}
\label{eq:distr2}
P(R_{g}^2)=\int\frac{\mathcal{D}\bold{r}}{Z_{0}}\exp\left[-\frac{d}{2Nb^2}\int\limits_{0}^{1}ds\dot{\bold{r}}^2(s)\right]
\delta\left(R_{g}^2-\frac{1}{2}\int\limits_{0}^{1}\!\!\!\int\limits_{0}^{1}ds_{1}ds_{2}\left(\bold{r}(s_{1})-\bold{r}(s_{2})\right)^2\right).
\end{equation}
It is worth noting that the integration in (\ref{eq:distr2}) is performed over the random vector-functions $\bold{r}(s)$ which describe the continuous curves in $d$-dimensional space.

Using the well known integral formula for the delta-function
\begin{equation}
\label{eq:delta}
\delta(x)=\int\limits_{-\infty}^{+\infty}\frac{d\xi}{2\pi}e^{-i\xi x},
\end{equation}
one can get the following representation for the PDF
\begin{equation}
\label{eq:distr2}
P(R_{g}^2)=\int\limits_{-\infty}^{+\infty}\frac{d\xi}{2\pi}K(\xi)e^{-i\xi R_{g}^2},
\end{equation}
where
\begin{equation}
\label{eq:char}
K(\xi)=\int \frac{\mathcal{D}\bold{r}}{Z_{0}}\exp\left[-\frac{d}{2Nb^2}\int\limits_{0}^{1}ds\dot{\bold{r}}^2(s)+\frac{i\xi}{2}\int\limits_{0}^{1}\!\!\!\int\limits_{0}^{1}ds_{1}ds_{2}\left(\bold{r}(s_{1})-\bold{r}(s_{2})\right)^2\right]
\end{equation}
is the standard characteristic function \cite{Gnedenko}.
Thus the initial problem is reduced to the couple of intermediate tasks. Firstly, we shall calculate the characteristic function $K(\xi)$ and secondly estimate the integral (\ref{eq:distr2}). As is well known, the characteristic function $K(\xi)$ is the generating function of moments
\begin{equation}
\mu_{k}=\int\limits_{-\infty}^{+\infty}P(u)u^{k}du
\end{equation}
\cite{Gnedenko}, so the following relation is valid
\begin{equation}
K(\xi)=\sum\limits_{k=0}^{\infty}\frac{\mu_{k}}{k!}(i\xi)^{k}.
\end{equation}
The cumulants $\chi_{k}$ can be evaluated from the following power series
\begin{equation}
\label{eq:cum}
\ln{K(\xi)}=\sum\limits_{k=1}^{\infty}\frac{\chi _{k}}{k!}(i\xi)^{k},
\end{equation}
It should be noted that cumulants $\chi _{k}$ define via moments $\mu_{k}$ by the relations $\chi _{1}=\mu_{1}$, $\chi _{2}=\mu_{2}-\mu_{1}^2$, etc \cite{Gnedenko,Kubo}.

\subsection{Characteristic function calculation}
After some algebra in integrand of (\ref{eq:char}) one can obtain the following expression
\begin{equation}
K(\xi)=\int\frac{\mathcal{D}\bold{r}}{Z_{0}}\exp\left[-\frac{d}{2Nb^2}\int\limits_{0}^{1}ds(\dot{\bold{r}}^2(s)-\omega^2\bold{r}^2(s))-i\xi\left(\int\limits_{0}^{1}ds\bold{r}(s)\right)^2\right],
\end{equation}
where $\omega^2(\xi)=2Nb^2i\xi/d$.
Using the well known integral relation
\begin{equation}
\int\limits_{-\infty}^{+\infty}dx e^{-ax^2+ibx}=\sqrt{\frac{\pi}{a}}e^{-\frac{b^2}{4a}},
\end{equation}
we arrive at
\begin{equation}
K(\xi)=\left(\frac{1}{4\pi i\xi}\right)^{d/2}\int d\bold{x}\exp\left[\frac{i\bold{x}^2}{4\xi}\right]\nonumber
\end{equation}
\begin{equation}
\times\int\frac{\mathcal{D}\bold{r}}{Z_{0}}\exp\left[-\frac{d}{2Nb^2}\int\limits_{0}^{1}ds(\dot{\bold{r}}^2(s)-\omega^2\bold{r}^2(s))+i\bold{x}\int\limits_{0}^{1}ds\bold{r}(s)\right].
\end{equation}

Thus, the task of characteristic function calculation reduces to the calculation of Gaussian path integral. In order to calculate the above Gaussian path integral, we have to specify the boundary condition for the random functions $\bold{r}(s)$. Fixing one of its ends at the origin, i.e., assuming $\bold{r}(0)=0$ and extracting the explicit integration over the second end of the chain $\bold{r}(1)=\bold{R}$, we obtain
\begin{equation}
\label{eq:char2}
K(\xi)=\left(\frac{1}{4\pi i\xi}\right)^{d/2}\left(\frac{d}{2\pi Nb^2}\right)^{d/2}\int d\bold{x}\exp\left[\frac{i\bold{x}^2}{4\xi}\right]
\int d\bold{R}\int\limits_{\bold{r}(0)=0}^{\bold{r}(1)=\bold{R}}\frac{\mathcal{D}\bold{r}}{Z_{1}}\exp\left[-S[\bold{r}]\right],
\end{equation}
where the following short-hand notation for the functional
\begin{equation}
S[\bold{r}]=\frac{d}{2Nb^2}\int\limits_{0}^{1}ds\left(\dot{\bold{r}}^2(s)-\omega^2\bold{r}^2(s)\right)-i\bold{x}\int\limits_{0}^{1}ds\bold{r}(s).
\end{equation}
is introduced;
\begin{equation}
Z_{1}=\int\limits_{\bold{\rho}(0)=0}^{\bold{\rho}(1)=0}\mathcal{D}\bold{\rho} \exp\left[-\frac{d}{2Nb^2}\int\limits_{0}^{1}ds\dot{\bold{\rho}}^{2}(s)\right]
\end{equation}
is the new normalization constant.
Now let us calculate the Gaussian path integral over $\bold{r}(s)$ in (\ref{eq:char2}) by means of the saddle-point
method which in the case of Gaussian path integrals give an exact result \cite{Zinn-Justin}. According to the saddle-point
method we assume that $\bold{r}(s)=\bold{v}_{SP}(s)+\bold{\rho}(s)$, where the function $\bold{v}_{SP}(s)$ satisfies the standard Euler-Lagrange equation
\begin{equation}
\frac{\delta S[\bold{v}_{SP}]}{\delta\bold{v}_{SP}(s)}=\frac{d}{Nb^2}\left(\ddot{\bold{v}}_{SP}(s)+\omega^2\bold{v}_{SP}(s)\right)+i\bold{x}=0\nonumber
\end{equation}
or
\begin{equation}
\label{eq:diff}
\ddot{\bold{v}}_{SP}(s)+\omega^2\bold{v}_{SP}(s)=-\frac{iNb^2\bold{x}}{d}
\end{equation}
with the boundary conditions $\bold{v}_{SP}(0)=0$ and $\bold{v}_{SP}(1)=\bold{R}$; $\bold{\rho}(s)$ is the function which describes the random fluctuations near the saddle-point \cite{Zinn-Justin} and satisfies the zeroth boundary conditions $\bold{\rho}(0)=\bold{\rho}(1)=0$.
Therefore, we obtain
\begin{equation}
K(\xi)=\left(\frac{1}{4\pi i\xi}\right)^{d/2}\left(\frac{d}{2\pi Nb^2}\right)^{d/2}G(\xi)\int d\bold{x}e^{\frac{i\bold{x}^2}{4\xi}}\int d\bold{R}e^{-S[\bold{v}_{SP}]},
\end{equation}
where the function $G(\xi)$ can be represented as the following Gaussian path integral
\begin{equation}
\label{eq:G}
G(\xi)=\int\limits_{\bold{\rho}(0)=0}^{\bold{\rho}(1)=0}\frac{\mathcal{D}\bold{\rho}}{Z_{1}}\exp\left[-\frac{d}{2Nb^2}\int\limits_{0}^{1}ds\left(\dot{\bold{\rho}}^{2}(s)-\omega^2\bold{\rho}^2(s)\right)\right].
\end{equation}
To calculate the function $G(\xi)$, we expand the random functions $\bold{\rho}(s)$ in the Fourier-series $\bold{\rho}(s)=\sqrt{2}\sum\limits_{n=1}^{\infty}\bold{\rho}_{n}\sin(\pi ns)$
and move to the integration in the path integral (\ref{eq:G}) over their Fourier-components $\bold{\rho}_{n}$ that yields
\begin{equation}
\label{eq:G2}
G(\xi)=\frac{\int \prod\limits_{n=1}^{\infty}d\bold{\rho}_{n}\exp\left[-\frac{d}{2Nb^2}\sum\limits_{n=1}^{\infty}\left(\pi^2n^2-\omega^2\right)
\bold{\rho}_{n}^2\right]}{\int \prod\limits_{n=1}^{\infty}d\bold{\rho}_{n}\exp\left[-\frac{d}{2Nb^2}\sum\limits_{n=1}^{\infty}\pi^2n^2\bold{\rho}_{n}^2\right]}
=\left(\prod\limits_{n=1}^{\infty}\frac{\pi^2n^2}{\pi^2n^2-\omega^2}\right)^{d/2}=\left(\frac{\omega}{\sin\omega}\right)^{d/2}.
\end{equation}
Further, solving the linear differential equation (\ref{eq:diff}), we arrive at
\begin{equation}
\bold{v}_{SP}(s)=-\frac{i\bold{x}Nb^2}{d\omega^2}\left(1-\cos(\omega s)-\frac{\sin(\omega s)\left(1-\cos{\omega}\right)}{\sin{\omega}}\right)+\frac{\bold{R}\sin(\omega s)}{\sin{\omega}}.
\end{equation}
The functional $S[\bold{v}_{SP}]$ takes the following form
\begin{equation}
S[\bold{v}_{SP}]=\frac{d}{2Nb^2}\int_{0}^{1}ds\left(\dot{\bold{v}}_{SP}^2(s)-\omega^2\bold{v}_{SP}^2(s)\right)-i\bold{x}\int_{0}^{1}ds\bold{v}_{SP}(s)=\nonumber
\end{equation}
\begin{equation}
=\frac{d\bold{v}_{SP}(1)\dot{\bold{v}}_{SP}(1)}{2Nb^2}-\frac{i\bold{x}}{2}\int_{0}^{1}ds\bold{v}_{SP}(s)=\frac{dR^2\omega}{2Nb^2\tan{\omega}}-i\bold{R}\bold{x}\frac{\tan{\frac{\omega}{2}}}{\omega}-
\frac{i\bold{x}^2}{2\xi}\frac{\tan{\frac{\omega}{2}}}{\omega}+\frac{i\bold{x}^2}{4\xi}.
\end{equation}
Hence, we get
\begin{equation}
K(\xi)=\left(\frac{1}{4\pi i\xi}\right)^{d/2}\left(\frac{d\omega}{2\pi Nb^2\sin{\omega}}\right)^{d/2}\times\nonumber
\end{equation}
\begin{equation}
\times\int d\bold{x}\int d\bold{R}\exp\left[-\frac{dR^2\omega}{2Nb^2\tan{\omega}}+\frac{i\bold{x}^2}{2\xi}\frac{\tan{\frac{\omega}{2}}}{\omega}+i\bold{R}\bold{x}\frac{\tan{\frac{\omega}{2}}}{\omega}\right]=
\label{eq:final1}
\left(\frac{\omega}{\sin{\omega}}\right)^{d/2}.
\end{equation}
We would like to stress that expression (\ref{eq:final1}) at $d=3$ is a result of Fixman \cite{Fixman} which was obtained within the Wang-Uhlenbeck method \cite{Yamakawa}.
The power series
$$\ln{\frac{\omega}{\sin{\omega}}}=\sum\limits_{k=1}^{\infty}\frac{2^{2k-1}B_{2k-1}}{k(2k)!}\omega^{2k}$$
together with the relation (\ref{eq:cum}) gives the following exact relations for the cumulant
\begin{equation}
\chi_{k}=\frac{2^{2k-1}(k-1)!(Nb^2)^{k}}{(2k)!}\left(\frac{2}{d}\right)^{k-1}B_{2k-1},
\end{equation}
where the Bernoulli numbers $B_{k}$ are introduced. Hence, we obtain the moments
\begin{equation}
\mu_{1}=\chi_{1}=\left<R_{g}^2\right>=\frac{Nb^2}{6},~\mu_{2}=\chi_{1}^2+\chi_{2}=\frac{4+5d}{180d}(Nb^2)^2,...
\end{equation}
As is seen, the first moment is a well known relation for the mean-square radius of gyration of the linear Gaussian polymer chain \cite{Yamakawa,Khohlov_book}.

\subsection{Probability distribution function calculation}
Let us estimate the PDF of the radius of gyration. Using the characteristic function expression (\ref{eq:final1}) and substituting it into the expression (\ref{eq:distr2}), one can obtain
\begin{equation}
\label{eq:P}
P(R_{g}^2)=\int\limits_{-\infty}^{\infty}\frac{d\xi}{2\pi}\exp\left[-W(\xi)\right],
\end{equation}
where the short-hand notation
\begin{equation}
\label{eq:W}
W(\xi)=\frac{\alpha^2\omega^2(\xi)d}{12}-\frac{d}{2}\ln{\frac{\omega(\xi)}{\sin{\omega(\xi)}}},
\end{equation}
and the notation for the expansion factor $\alpha=R_{g}/\left<R_{g}^2\right>^{1/2}$ are introduced. We shall estimate the integral (\ref{eq:P}) by the steepest-descent method. We would like to stress that further calculations will be presented quite similar to that are in the ref. \cite{Fixman} for the case of $d=3$.

The saddle-point equation has a form
\begin{equation}
\label{eq:saddle}
W^{\prime}(\xi_0)=\frac{d}{2\xi_0}\left(\frac{\alpha^2\omega^2(\xi_0)}{6}-\frac{1}{2}\left(1-\frac{\omega(\xi_0)}{\tan{\omega(\xi_0)}}\right)\right)=0.
\end{equation}

Thus in the vicinity of saddle-point $\xi_{0}$ we get
\begin{equation}
W(\xi)=W(\xi_{0})+\frac{1}{2}W^{\prime\prime}(\xi_{0})\left(\xi-\xi_{0}\right)^2+...
\end{equation}
We would like to stress that the contour of integration must be deformed along the line of steepest descent \cite{Zinn-Justin}. The second derivative $W^{\prime\prime}(\xi)$ has a following form
\begin{equation}
\label{eq:second_deriv}
W^{\prime\prime}(\xi)=\frac{d}{8\xi^2}\left(2-\frac{\omega}{\tan{\omega}}-\frac{\omega^2}{\sin^2{\omega}}\right).
\end{equation}
The saddle-point equation can be easily solved in two limiting cases which are of interest for physical applications, namely, for $\alpha\gg 1$ and $\alpha\ll 1$.

In the case of large radius of gyration ($\alpha\gg 1$) we obtain the evaluation for the saddle-point $\omega(\xi_0)=\omega_{0}\simeq\pi-3/\pi\alpha^2$ which yields
\begin{equation}
P(R_{g}^2)\simeq e^{\frac{d}{2}-\frac{\pi^2 d}{12}\alpha^2-\frac{d}{2}\ln\frac{3}{\pi^2\alpha^2}}\int\limits_{-\infty}^{\infty}\frac{d\xi}{2\pi}\exp\left[-\frac{\pi^2 N^2 b^4\alpha^4}{36d}(\xi-\xi_{0})^2\right].
\end{equation}
Further, taking the Gaussian integral in (\ref{eq:P2}), we obtain eventually
\begin{equation}
\label{eq:P2}
P(R_{g}^2)\simeq\frac{e^{d/2}3^{(2-d)/2}\pi^{d-3/2}\sqrt{d}}{Nb^2}\alpha^{d-2}\exp\left[-\frac{\pi^2d}{12}\alpha^2\right].
\end{equation}
It should be noted that at $d=3$ we obtain the expression which was first obtained by Fixman within the Wang-Uhlenbeck method \cite{Fixman}. Moreover, the expression (\ref{eq:P2}) was first obtained  recently as an intermediate result in the work \cite{Brilliantov} and used in works \cite{Brilliantov,Seidel}.

In the case of small radius of gyration, when $\alpha\ll 1$, we obtain the saddle-point $\omega(\xi_{0})=\omega_{0}\simeq 3i /\alpha^2$. The probability distribution function in this case can be evaluated by the steepest-descent method in the following way
\begin{equation}
\label{eq:P3}
P(R_{g}^2)\simeq e^{-\frac{3d}{4\alpha^2}-\frac{d}{2}\ln{\frac{\alpha^2}{6}}}\int\limits_{-\infty}^{\infty}\frac{d\xi}{2\pi}\exp\left[-\frac{N^2 b^4\alpha^6}{108d}(\xi-\xi_{0})^2\right].
\end{equation}
After the calculation of Gaussian integral and some elementary algebraic transformations we arrive at
\begin{equation}
P(R_{g}^2)\simeq \sqrt{\frac{27}{\pi}}6^{d/2} \alpha^{-3-d}\frac{e^{-\frac{3d}{4\alpha^2}}}{Nb^2}.
\end{equation}
In the case of three-dimensional space ($d=3$) we obtain the relation (up to a numerical prefactor) which was obtained by Fixman in the work \cite{Fixman}.

Putting together the above results, we obtain eventually the following limiting laws
\begin{equation}
\label{eq:asymp1}
P(R_{g}^2)\simeq
 \begin{cases}
\frac{e^{d/2}3^{(2-d)/2}\pi^{d-3/2}\sqrt{d}}{Nb^2}\alpha^{d-2}\exp\left[-\frac{\pi^2d}{12}\alpha^2\right], &\text{if $\alpha\gg 1$}\\
\left(\frac{27}{\pi}\right)^{1/2}\frac{6^{d/2}\alpha^{-3-d}}{Nb^2}\exp\left[-\frac{3d}{4\alpha^2}\right], &\text{if $\alpha\ll 1$}.
 \end{cases}
\end{equation}

It should be noted that the limiting laws (\ref{eq:asymp1}) for $d=3$ are widely used in the different applications of polymer physics \cite{Grosberg1,Brilliantov2,Budkov1,Budkov2,Budkov3,Budkov4,Budkov5}.

\section{Example II: PDF of potential energy of Gaussian polymer chain in d-dimensional parabolic well}
The second example of the application of our method which has rather academic interest is calculation of the PDF of potential energy of the Gaussian polymer chain in $d$-dimensional parabolic well. We consider only the case of symmetric parabolic well, since the generalization on the asymmetric case is trivial.
\subsection{Statement of problem}
We start from the PDF of potential energy $\epsilon$ of the Gaussian polymer chain in $d$-dimensional symmetric parabolic well, written in the form of the following path integral:
\begin{equation}
\label{eq:distr_en}
P(\epsilon)=\int\frac{\mathcal{D}\bold{r}}{Z_{0}}\exp\left[-\frac{d}{2b^2}\int\limits_{0}^{N}d\tau\dot{\bold{r}}^2(\tau)\right]
\delta\left(\epsilon-\frac{\kappa}{2}\int\limits_{0}^{N}d\tau\bold{r}^2(\tau)\right),
\end{equation}
where $\kappa$ is a coefficient of stiffness and
\begin{equation}
Z_{0}=\int \mathcal{D}\bold{r}\exp\left[-\frac{d}{2b^2}\int\limits_{0}^{N}d\tau\dot{\bold{r}}^2(\tau)\right]
\end{equation}
is the normalization constant. We also assume that one of the ends of the polymer chain is fixed at the origin, i.e., $\bold{r}(0)=0$. In addition, we assume that the center of the parabolic well is also located at the origin.
Using the substitution $\tau=sN$ as well as in the previous section, where $0\leq s\leq 1$, the PDF can be rewritten as follows
\begin{equation}
\label{eq:distr_en2}
P(\epsilon)=\int\frac{\mathcal{D}\bold{r}}{Z_{0}}\exp\left[-\frac{d}{2Nb^2}\int\limits_{0}^{1}ds\dot{\bold{r}}^2(s)\right]
\delta\left(\epsilon-\frac{\kappa N}{2}\int\limits_{0}^{1}ds\bold{r}^2(s)\right).
\end{equation}
Further, applying the formula (\ref{eq:delta}) for the delta function in integrand of (\ref{eq:distr_en2}), we arrive at the following Fourier representation of the PDF:
\begin{equation}
\label{eq:distr_en3}
P(\epsilon)=\int\limits_{-\infty}^{+\infty}\frac{d\xi}{2\pi}K(\xi)e^{-i\xi \epsilon},
\end{equation}
where the characteristic function
\begin{equation}
\label{eq:char_en}
K(\xi)=\int \frac{\mathcal{D}\bold{r}}{Z_{0}}\exp\left[-\frac{d}{2Nb^2}\int\limits_{0}^{1}ds\dot{\bold{r}}^2(s)+\frac{i\xi\kappa N}{2}\int\limits_{0}^{1}ds\bold{r}^2(s)\right]
\end{equation}
is introduced.
Now we have to calculate the characteristic function as a simple Gaussian path integral and then estimate the PDF. It should be noted that in contrast to the WU method which might be also applied to this problem and would be related to cumbersome calculations, the present approach reduces the characteristic function calculation to the simple calculation of standard Gaussian path integral \cite{Zinn-Justin}.

\subsection{Characteristic function calculation}
The characteristic function (\ref{eq:char_en}) can be rewritten in the following form:
\begin{equation}
\label{eq:char_en2}
K(\xi)=\int \frac{\mathcal{D}\bold{r}}{Z_{0}}\exp\left[-\frac{d}{2Nb^2}\int\limits_{0}^{1}ds\left(\dot{\bold{r}}^2(s)-\Omega^2\bold{r}^2(s)\right)\right],
\end{equation}
where $\Omega=\Omega(\xi)=\left(\frac{i\xi\kappa N^2b^2}{d}\right)^{1/2}$.

Further, extracting the integration over the second end of the polymer chain, one can get
\begin{equation}
K(\xi)=\int d\bold{R}\int\limits_{\bold{r}(0)=0}^{\bold{r}(1)=\bold{R}}\frac{\mathcal{D}\bold{r}}{Z_{0}}\exp\left[-S[\bold{r}]\right],
\end{equation}
where the functional
\begin{equation}
S[\bold{r}]=\frac{d}{2Nb^2}\int\limits_{0}^{1}ds\left(\dot{\bold{r}}^2(s)-\Omega^2\bold{r}^2(s)\right)
\end{equation}
is introduced.
As well as in the previous section we calculate the Gaussian path integral by the saddle-point method. We represent the random function $\bold{r}(s)$ as a sum:
\begin{equation}
\bold{r}(s)=\bold{v}_{SP}(s)+\bold{\rho}(s).
\end{equation}
Here, the function $\bold{v}_{SP}(s)$ satisfies the saddle-point equation
\begin{equation}
\frac{\delta S[\bold{v}_{SP}]}{\delta\bold{v}_{SP}(s)}=\frac{d}{Nb^2}\left(\ddot{\bold{v}}_{SP}(s)+\Omega^2\bold{v}_{SP}(s)\right)=0\nonumber,
\end{equation}
or
\begin{equation}
\label{eq:diff2}
\ddot{\bold{v}}_{SP}(s)+\Omega^2\bold{v}_{SP}(s)=0,
\end{equation}
and the following boundary conditions $\bold{v}_{SP}(0)=0$ and $\bold{v}_{SP}(1)=\bold{R}$.
The random vector-function $\bold{\rho}(s)$ satisfies the zero boundary conditions $\bold{\rho}(0)=\bold{\rho}(1)=0$ and describes the random fluctuations near the extremal $\bold{v}_{SP}(s)$.
Solution of the equation (\ref{eq:diff2}) yields
\begin{equation}
\bold{v}_{SP}(s)=\frac{\bold{R}\sin(\Omega s)}{\sin\Omega}.
\end{equation}
Further, performing the transformations as well as in the previous section, we arrive at
\begin{equation}
K(\xi)=\left(\frac{d}{2\pi Nb^2}\right)^{d/2}G(\xi)\int d\bold{R}\exp\left[-S[\bold{v}_{SP}]\right],
\end{equation}
where
\begin{equation}
G(\xi)=\left(\frac{\Omega}{\sin{\Omega}}\right)^{d/2},
\end{equation}
and
\begin{equation}
S[\bold{v}_{SP}]=\frac{d\bold{v}_{SP}(1)\dot{\bold{v}}_{SP}(1)}{2Nb^2}=\frac{d\Omega R^2}{2Nb^2\tan{\Omega}}.
\end{equation}

Calculating the Gaussian integral over $\bold{R}$ and performing some algebraic transformations, we obtain eventually
\begin{equation}
\label{eq:K}
K(\xi)=\left(\frac{1}{\cos{\Omega(\xi)}}\right)^{d/2}.
\end{equation}

The formula (\ref{eq:K}) allows us to calculate the cumulants and moments of the distribution. The power series
\begin{equation}
\ln(\cos{\Omega})=\sum\limits_{n=1}^{\infty}\frac{(-1)^{n+1}B_{2n}2^{2n-1}(1-2^{2n})}{(2n)!n}\Omega^{2n}
\end{equation}
together with the relation (\ref{eq:cum}) gives the following relation for the cumulants
\begin{equation}
\chi_{k}=\frac{(-1)^{k}(k-1)!B_{2k}2^{2k-2}(1-2^{2k})}{(2k)!d^{k-1}}(\kappa N^2b^2)^{k}.
\end{equation}
Therefore, we obtain the moments
\begin{equation}
\mu_{1}=\chi_{1}=\left<\epsilon\right>=\frac{\kappa N^2b^2}{4},~\mu_{2}=\chi_2+\mu_1^2=\left<\epsilon^2\right>=\frac{3d+4}{48d}\left(\kappa N^2b^2\right)^2,..
\end{equation}
It is interesting to note that average potential energy of the polymer chain (as well as the mean-square radius of gyration in previous section) does not depend on the dimension of space $d$. It is also worth noting that the mean-potential energy of the polymer chain can be rewritten as $\left<\epsilon\right>=3N\kappa\left<R_{g}^{2}\right>/2$. It means that each monomer in the symmetric parabolic well is displaced in average from the origin onto the distance $\sqrt{3\left<R_{g}^{2}\right>}$.

\subsection{Probability distribution function calculation}
Now let us estimate the PDF by the standard steepest-descent method. Using the characteristic function expression (\ref{eq:K}) and substituting it into the expression (\ref{eq:distr_en3}), one can obtain
\begin{equation}
\label{eq:P_en}
P(\varepsilon)=\int\limits_{-\infty}^{\infty}\frac{d\xi}{2\pi}\exp\left[-W(\xi)\right],
\end{equation}
where
\begin{equation}
W(\xi)=\frac{d}{2}\left(\frac{\Omega^2(\xi)\lambda}{2}+\ln{\cos{\Omega(\xi)}}\right)
\end{equation}
and the dimensionless energy $\lambda=\epsilon/\left<\epsilon\right>$ is also introduced.

The saddle-point equation has a form:
\begin{equation}
W^{\prime}(\xi_{0})=\frac{d\Omega^2(\xi_{0})}{4\xi_{0}}\left(\lambda-\frac{\tan{\Omega(\xi_{0})}}{\Omega(\xi_{0})}\right)=0.
\end{equation}

Thus, in the vicinity of the saddle-point we have the following expansion:
\begin{equation}
W(\xi)=W(\xi_{0})+\frac{1}{2}W^{\prime\prime}(\xi_{0})(\xi-\xi_{0})^2+..,
\end{equation}
where the second derivative is
\begin{equation}
W^{\prime\prime}(\xi_{0})=\frac{d\Omega(\xi_{0})}{8\xi_{0}^2}\left(\tan\Omega(\xi_{0})-\frac{\Omega(\xi_{0})}{\cos^2{\Omega(\xi_{0})}}\right).
\end{equation}
The saddle-point equation can be easily solved in two limiting cases: when $\lambda\gg 1$ and $\lambda\ll 1$. In the case when $\lambda\gg 1$ we obtain the following saddle-point $\Omega_{0}=\Omega(\xi_{0})\simeq \pi/2 -2/{\pi\lambda}$, so that
\begin{equation}
W(\xi_{0})\simeq\frac{d}{2}\left(\frac{\pi^2\lambda}{8}-2+\ln{\frac{2}{\pi\lambda}}\right),
\end{equation}
and
\begin{equation}
W^{\prime\prime}(\xi_{0})\simeq \frac{\kappa^2N^4b^4}{8d}\lambda^2.
\end{equation}
Taking the Gaussian integral, we arrive at
\begin{equation}
\label{eq:P_lar_en}
P(\epsilon)\simeq \frac{\pi^{(d-1)/2}e^{d}\sqrt{d}\lambda^{d/2-1}}{2^{d/2-1}\kappa N^2 b^2}\exp\left[-\frac{\pi^2\lambda}{16}\right].
\end{equation}

As is seen from eq. (\ref{eq:P_lar_en}), at the region of large potential energy of the polymer chain the PDF is reminiscent the Boltzmann distribution function with effective temperature $T=(16/\pi^2)\left<\epsilon\right>$.

In the opposite case $\lambda\ll 1$ we have the saddle-point $\Omega_{0}=\Omega(\xi_{0})\simeq i/\lambda$, so that
\begin{equation}
W(\xi_{0})\simeq \frac{d}{4\lambda}-\frac{d}{2}\ln{2},
\end{equation}
and
\begin{equation}
W^{\prime\prime}(\xi_{0})\simeq \frac{\kappa^2N^4b^4\lambda^3}{8d}.
\end{equation}

The calculation of Gaussian integral yields in this case:
\begin{equation}
\label{eq:P_sm_en}
P(\epsilon)\simeq \frac{2^{d/2+1}}{\kappa N^2b^2}\left(\frac{d}{\pi\lambda^3}\right)^{1/2}\exp\left[-\frac{d}{4\lambda}\right].
\end{equation}
As is seen from eq. (\ref{eq:P_sm_en}), the PDF must converge very fast to zero at the small potential energy of polymer chain.

Collecting together eqs. (\ref{eq:P_lar_en}) and (\ref{eq:P_sm_en}), we arrive at
\begin{equation}
\label{eq:asymp1}
P(\epsilon)\simeq
 \begin{cases}
\frac{\pi^{(d-1)/2}e^{d}\sqrt{d}\lambda^{d/2-1}}{2^{d/2-1}\kappa N^2 b^2}\exp\left[-\frac{\pi^2\lambda}{16}\right], &\text{if $\lambda\gg 1$}\\
\frac{2^{d/2+1}}{\kappa N^2b^2}\left(\frac{d}{\pi\lambda^3}\right)^{1/2}\exp\left[-\frac{d}{4\lambda}\right], &\text{if $\lambda\ll 1$}.
 \end{cases}
\end{equation}

\section{Concluding remarks}
In this work we have developed a new approach based on the path integrals formalism for the calculation of the probability distribution functions of the quadratic quantities of the Gaussian polymer chain in $d$-dimensional space. We demonstrate the applicability of our method to the calculation of probability distribution functions of the radius of gyration and potential energy in the symmetric parabolic well for the Gaussian polymer chain. In both cases we have obtained the exact relations for characteristic functions and cumulants and calculated the few first moments. We have established the limiting laws for the probability distribution functions at the large and small values of the corresponding variables. We have generalized the classical Fixman's result for the probability distribution function of radius of gyration of the Gaussian polymer chain in space of arbitrary dimension $d$. We have showed the obvious advantage of our method compared to the standard Wang-Uhlenbeck method, calculating the probability distribution function of the potential energy of Gaussian polymer chain in $d$-dimensional parabolic well. The additional advantage of proposed method is that it can be easily applied to the calculation of the probability distribution functions of the quadratic quantities of the more complex objects, such as the ideal diblock-copolymer and worm-like polymer chain. These results will be published elsewhere \cite{Budkov6, Budkov7}.

\section{Acknoledgements}
YAB thanks N.V. Brilliantov for motivating discussions. We thank Reviewer for valuable comments that helped us to improve this work.

\end{document}